\documentclass[letterpaper]{article}
\usepackage{aaai2026}
\usepackage{times}
\usepackage{helvet}
\usepackage{courier}
\usepackage[hyphens]{url}
\usepackage{graphicx}
\usepackage{natbib}
\usepackage{caption}
\nocopyright %

\usepackage{tikz}
\usetikzlibrary{shapes, arrows.meta, positioning, calc}

\usepackage{booktabs}
\usepackage{array} %

\usepackage{todonotes}

\usepackage{cleveref}

\definecolor{teal}{rgb}{0.0, 0.5, 0.5}
\definecolor{forestgreen}{HTML}{228B22}

\def\extendedversion{}

\title{Agentic AI: User Empowerment or Foreclosure?}

\author{
    David Gamba\textsuperscript{\rm 1},
    Daniel M. Romero\textsuperscript{\rm 1},
    Grant Schoenebeck\textsuperscript{\rm 1}
}
\affiliations{
    \textsuperscript{\rm 1}University of Michigan\\
    Ann Arbor, MI, USA\\
    gamba@umich.edu, drom@umich.edu, schoeneb@umich.edu
}

\begin{document}

\maketitle

\begin{abstract}
Agentic AI promises systems that can act on users' behalf, from filtering content to negotiating prices to selecting services. Whether it will empower users is an open question, and one that depends on more than the technology. We conduct a comparative case analysis of four earlier, more mature domains in which similar forms of agency emerged: browser-based ad blockers, platform recommender systems, financial robo-advisors, and email spam filtering. Across the cases, questions about whose interests agents would serve were resolved through technical arrangements: API choices, protocol governance, industry standards, and default configurations. Beyond their technical form, these were political decisions. We identify this settling of contestable questions in a technical form as depoliticization, a concept from political theory, here at work in technological systems. Its most consequential effect is that individual outcomes and collective contestation capacity can move in opposite directions: spam inbox quality improved substantially while the organized capacity to contest spam governance collapsed. Where intermediary institutions sustained formal channels for challenge, user-aligned agency proved more durable; where proprietary infrastructure and closed standard-setting absorbed contestation, the material basis for user-aligned alternatives was dismantled, and the loss proved hard to reverse. Applying this lens to agentic AI, we find a similar pattern forming: governance is consolidating around the Model Context Protocol and the Agentic AI Foundation, an industry-governed venue already deciding what agents will be able to do. Unlike in the completed trajectories, these decisions have not yet hardened, and remain open to challenge by users and the public.
\end{abstract}

\section{Introduction}
\label{sec:intro}

In 2019, Google announced a redesign of the extension API for its Chrome browser, later shipped as Manifest V3. No ad blocker was banned. No filtering rule was invalidated. Instead, the operations available to browser extensions were redefined: the flexible \texttt{webRequest} API, which allowed extensions to intercept and modify network requests in real time, was replaced with the more constrained \texttt{declarativeNetRequest} API, which requires extensions to declare filtering rules in advance and limits how many rules can be active simultaneously. The practical effect was to substantially reduce the capability of content‑blocking extensions~\citep{cyphersGooglesManifestV32021}. A contestable question (what should users be able to block in their browsers, and who decides) was settled by being embedded in browser architecture. What makes this a political act is that it was a choice among alternatives, made by an actor with interests in the outcome.

Agentic AI is a new technology that, like ad blockers,  promises to act on users' behalf, though in more powerful ways. Could an ordinary user, for example, direct an agent to build and maintain their own browser based on Chromium that blocks ads following Firefox's implementation? Or, less ambitiously, could they build browser extensions that reorder or filter a social media feed or search results according to their personal instructions? Of course, the platforms may attempt to block such behavior.  In the ensuing cat-and-mouse game, who wins?  

To better understand the potential outcomes of this dynamic, we look to predecessors of agentic AI, systems that also promised to act in the interests of the user: ad blockers, recommender systems, robo-advisors, and spam filters. Each initially opened space for user-aligned action, and each has attracted a substantial critical literature. We perform a comparative analysis of these trajectories in detail to trace how that space shifted over time.

The focus of our study is \emph{user empowerment}, by which we mean the capacity to meaningfully influence how an agent acts in ways that align with the user's own interests.
That capacity is exercised individually but preserved, or lost, collectively. Our focus is on what kinds of agents are even possible in a domain: which optimization targets, data sources, and actions on behalf of users are available to any agent, regardless of which firm builds it. We call this the \emph{boundaries of possible agency}.  Whether an agent may build an ad-blocking browser, and whether an extension may reorder a feed, are questions about where those boundaries sit.

Comparing the trajectories reveals a structural tendency: \emph{depoliticization}, the progressive removal of contestable choices from open challenge by embedding them in technical and institutional forms~\citep{burnhamNewLabourPolitics2001, flindersDepoliticisationPrinciplesTactics2006}. These trajectories admit competing explanations, and for particular moments strong ones are available. The turn in spam filtering reads naturally as an arms race won by better classifiers, and Chrome's extension redesign was defended on security and performance grounds; market-selection and regulatory-capture readings fit other cases. We address these accounts where they bear and argue that none, though each may explain a moment, suffices to explain what the comparison reveals. Across four domains and more than two decades, the cases differ in whether organized collective contestation ever formed and, where it did, whether it survived.

Building on the depoliticization tendency the cases reveal, we advance three contributions. First, using historical trajectories of user‑facing technologies, we show that improvements in individual user experience and collective contestation capacity can move in opposite directions, and we specify conditions under which that divergence occurs.

Second, we develop a constitutive politics framework that locates where user‑aligned agency is foreclosed or sustained along three dimensions: infrastructural (what agents run on), epistemic (what they can know), and teleological (what they optimize for).  We argue that if user empowerment is to be durable, it must rest on the collective capacity to contest the choices that configure these dimensions.  Across the cases, that capacity was sustained where intermediary institutions provided adversarial governance, i.e., formal channels through which their decisions can be challenged (the SEC's rulemaking), and independent infrastructure (Firefox's extension API), and collapsed where proprietary infrastructure (Chrome's Manifest V3) and gated standard‑setting (the email-deliverability forum M3AAWG) foreclosed participation. When foreclosure accumulates across all three dimensions at once, the footholds from which contestation could be rebuilt are removed together, and rebuilding them is far harder than defending them would have been. 

Third, we apply the framework to agentic AI. It is now consolidating across a far broader set of domains than any of these predecessors, including commerce, information retrieval, and task automation, through a single protocol infrastructure: the Model Context Protocol. Anthropic released the protocol in late 2024, and major providers adopted it within twelve months~\citep{linuxFoundation2025AAIF}. Anthropic has since donated it to the Agentic AI Foundation at the Linux Foundation. There, a Governing Board composed of platform providers carries decision‑making authority. User organizations and public‑interest representatives have no formal standing. This configuration matches the foreclosure‑prone patterns the cases identify. But the specifications are young, and their choices are not yet locked into deployed infrastructure or user habit. We set out what preserving contestation would require while the architecture remains malleable.

\section{Constitutive Politics: A Lens}
\label{sec:framework}

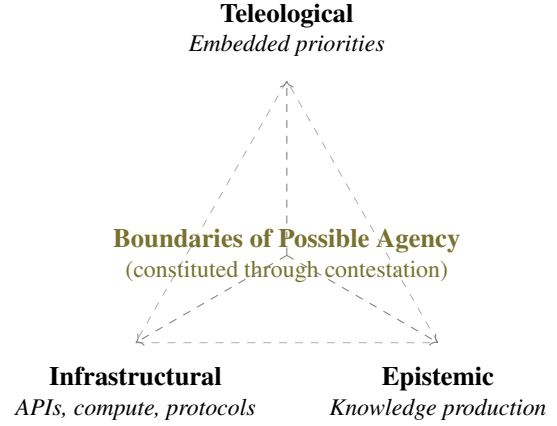
\begin{figure}[t]
\centering
\begin{tikzpicture}

    \coordinate (infra)     at (0, 0);
    \coordinate (epistemic) at (4, 0);
    \coordinate (value)     at (2, 3.464);

    \draw[dashed, gray!60] (infra) -- (epistemic) -- (value) -- cycle;

    \node[align=center, anchor=north] at ($(infra) + (0, -0.2)$) {
        \textbf{Infrastructural}\\
        {\small\emph{APIs, compute, protocols}}
    };

    \node[align=center, anchor=north] at ($(epistemic) + (0, -0.2)$) {
        \textbf{Epistemic}\\
        {\small\emph{Knowledge production}}
    };

    \node[align=center, anchor=south] at ($(value) + (0, 0.2)$) {
        \textbf{Teleological}\\
        {\small\emph{Embedded priorities}}
    };

    \coordinate (centroid) at (barycentric cs:infra=1,epistemic=1,value=1);
    \node[align=center, text=olive!70!black] at (centroid) {
        \textbf{Boundaries of Possible Agency}\\
        {\small(constituted through contestation)}
    };

    \draw[->, dashed, gray] (centroid) -- (infra);
    \draw[->, dashed, gray] (centroid) -- (epistemic);
    \draw[->, dashed, gray] (centroid) -- (value);

\end{tikzpicture}
\caption{Dimensions of constitutive politics that define the boundaries of agency. Struggles over these dimensions constitute what agents in a domain can be and whose interests they can serve.}
\label{fig:dimensions}
\end{figure}

To function as a computational system, any ``agent'' of the kind examined in this paper requires three things: infrastructure to run on (compute, APIs, protocols), data from which to reason, and an objective that defines what it is optimizing for~\citep{russellArtificialIntelligenceModern2010, wooldridgeIntelligentAgentsTheory1995}. This decomposition holds whether the system is a 1990s spam filter operating on fixed rules or a contemporary machine learning recommender. Each element is technically necessary; each is also a site of contestation (\cref{fig:dimensions}). Drawing on Winner's analysis of how artifacts embody political properties and Mouffe's concept of the political as the domain of open contestation, which persists only where institutions grant opposed parties standing as adversaries~\citep{winnerArtifactsPolitics1980, birchArtifactsHavePolitical2025, mouffeReturnPolitical2005, mouffePolitical2011}, we develop a \emph{constitutive politics} lens organized around these three sites. The lens asks what agents in a domain can be, what they can know, and whose interests they can serve. It then asks where those possibilities are set, by whom, and whether they could be otherwise.

Other lenses could of course be used. Principal-agent theory, AI alignment, and participatory design each offer tools for improving agent behavior given the conditions under which agents operate~\citep{duettingAlgorithmicContractTheory2024, sorensenRoadmapPluralisticAlignment2024, mullerParticipatoryDesign1993}. The constitutive politics lens asks how those conditions came to be configured. At each turning point the cases examine, multiple configurations were technically feasible; the question is what forces selected the one that was adopted. The comparison across the trajectories answers this. Each of the three dimensions indexes where those choices were made: in infrastructure, in what the system can know, in what it optimizes for, respectively.

\begin{description}
    \item[\emph{Infrastructural.}] The compute, APIs, protocols, browser engines, and data pipelines on which agents operate, along with the legal access rights that govern them. Control at any layer of this stack shapes what is possible at the layers above it, which is why a browser API redesign can redraw the boundaries for every blocking agent in a domain without targeting any one of them directly.

    \item[\emph{Epistemic.}] How agents produce knowledge about users and the world, and how that knowledge is represented. A community-maintained filter list and a proprietary behavioral classifier are both ways of knowing, yet they encode different answers to who can challenge the judgments embedded in the system and on what basis.

    \item[\emph{Teleological.}] The priorities embedded in agents' concrete configurations, including optimization targets, default settings, fee structures, and legal standards. For instance, when a platform encodes a particular metric as its recommendation objective, that is a political choice embedded in a loss function.
\end{description}

The cases that follow are histories of how these three dimensions were defined and contested over time, organized around the actors involved, their incentives, and the turning points that shaped each trajectory.

\section{Technological Trajectories}
\label{sec:cases}

The technologies examined here were each introduced, at different moments and in different domains, as tools that would act on behalf of users for specific tasks. Each has operated long enough to be studied extensively, and a substantial literature has documented their development. We selected the trajectories to span variation in how they developed: from sustained community governance to complete foreclosure, with intermediate cases of collapse and regulatory absorption. We present them as historical narratives organized around the actors involved and the major turning points that shaped each trajectory. How these turning points should be interpreted, and whether common patterns run across them, is a question we take up in subsequent sections.

\subsection{Browser-Based Ad Blockers}
\label{sec:case-adblockers}

In the mid-2000s browser extensions that intercept and filter advertising content emerged as a niche practice among technically adept users. By the 2010s they had become one of the most widely deployed user-side tools on the web. At their core, ad blockers combine two mechanisms: network-level request blocking, in which outgoing HTTP requests to known advertising domains are cancelled before loading, and cosmetic filtering, in which CSS rules hide ad-slot elements that survive request blocking~\citep{storeyFutureAdBlocking2017, nithyanandAdblockingCounterBlocking2016}. The technical details matter because both mechanisms depend on what browser extension APIs permit, and what they permit has itself become contested.

The early ad-blocking ecosystem was organized around shared filter lists. EasyList, maintained by a small group of volunteers and open to crowdsourced contributions, became the de facto standard, encoding rules about which domains, URL patterns, and page elements should be blocked~\citep{alrizahErrorsMisunderstandingsAttacks2019, snyderWhoFiltersFilters2020}. The list was publicly readable, forkable, and updated through a visible community process. Disputes about which ads should be blocked (including whether a given analytics domain belonged on a privacy list) were resolved through argument and revision within the community; anyone who disagreed with a decision could fork the list and maintain their own~\citep{snyderWhoFiltersFilters2020}. One estimate placed Adblock Plus alone at approximately 100 million users by the late 2010s~\citep{megaliDigitalPlatformsMembers2022}, and the economic stakes for publishers and advertisers rose with adoption. Web publishing had entered a sustained revenue crisis in the early 2010s, with advertising the primary source of income for news and content sites, and rising ad-block adoption was experienced by publishers as a direct threat to that revenue base~\citep{megaliDigitalPlatformsMembers2022, zhaoBeToughSoft2020}.

The first major institutional development came in 2011, when Eyeo GmbH, the company behind Adblock Plus, introduced the Acceptable Ads program~\citep{guptaBlockBlockerStudying2020}. Rather than blocking all advertising, Acceptable Ads defined criteria for “non-intrusive” formats (governing size, placement, animation, and labeling) and allowed compliant ads through by default, charging fees to larger advertisers seeking inclusion on the whitelist. Publishers and advertisers who objected to ad blocking had pursued legal challenges in German courts; those courts upheld both the act of blocking and the practice of charging for whitelisting as lawful market behavior, thereby consolidating the Acceptable Ads arrangement as a legally valid settlement of the dispute~\citep{nithyanandAdblockingCounterBlocking2016, guptaBlockBlockerStudying2020}. A contested question, whether and by whom advertising could be blocked, was restructured into a different one: which advertising formats met the technical criteria for the whitelist, and at what price.

Not all projects accepted this settlement. uBlock Origin, developed independently by Raymond Hill, rejected the Acceptable Ads framework and maintained community-curated, user-modifiable filtering without a whitelisting program~\citep{alrizahErrorsMisunderstandingsAttacks2019}. It offered users direct control over which lists to enable, the ability to add custom rules, and the option to allow or block individual elements on a per-site basis. That choice split the ecosystem: commercial blockers now operate within the Acceptable Ads framework, while projects like uBlock Origin maintain a meaningfully different relationship to the advertising industry.

Alongside the Acceptable Ads negotiation, a more confrontational technical contest developed between blockers and publisher-side counter-measures. Specialized anti-adblock vendors such as PageFair and BlockAdblock developed detection techniques that publishers deployed on roughly 6–7\% of high-traffic sites, using bait elements styled like ads, probes for missing ad-library objects, and behavioral network monitoring~\citep{nithyanandAdblockingCounterBlocking2016, guptaBlockBlockerStudying2020}. Filter-list maintainers responded with anti-anti-adblock rules and scriptlets targeting known detection patterns; researchers formalized the dynamic as a security-style arms race and proposed perceptual, stealth, and active blocking techniques as next-generation responses~\citep{storeyFutureAdBlocking2017, iqbalAdWarsRetrospective2017, alrizahErrorsMisunderstandingsAttacks2019}. Publishers experimented with both hard walls (requiring whitelisting to access content) and soft cooperative approaches (delivering “acceptable” less-intrusive ads via Acceptable Ads exchanges). Field experiments on Forbes showed that hard walls reduced engagement overall, particularly among less-loyal users, while soft cooperation preserved a broader audience and gradually converted some users to whitelisting~\citep{zhaoBeToughSoft2020}.

In 2017, the Coalition for Better Ads published standards defining ad formats it deemed unacceptable, based on consumer experience surveys, and delegated enforcement to member platforms, primarily through Chrome’s Ad Experience Report~\citep{megaliDigitalPlatformsMembers2022}. The Coalition included Google among its co-founders alongside major advertising trade associations and publishing companies; its membership did not include formal representation from filter-list communities or consumer or public-interest organizations, and its governance structure lacked mechanisms for adversarial challenge or an independent review process~\citep{blackburnDigitalPlatformsCircular2023}. Beginning in 2019, Chrome enforced Coalition standards as a built-in browser feature, automatically filtering ads that violated them for all users, regardless of whether they had installed any ad-blocking extension~\citep{megaliDigitalPlatformsMembers2022}. What counted as acceptable advertising, previously contested between publishers, advertisers, blocklist communities, and users, now fell largely within the discretion of Google, the largest seller of the advertising those standards govern.

The filter lists themselves had accumulated weaknesses that would later support the case for constraining them. Longitudinal analysis of EasyList found that roughly 90\% of resource-blocking rules were unused in typical browsing scenarios, contributing to computational overhead that mattered particularly on mobile devices. False positives could persist for weeks or months under the list’s highly centralized maintenance structure~\citep{snyderWhoFiltersFilters2020, alrizahErrorsMisunderstandingsAttacks2019}. Against this background, the most recent significant development has been the transition from Chrome’s Manifest V2 extension API to Manifest V3, announced in 2019 and progressively enforced in subsequent years. The older API permitted extensions to use a flexible \texttt{webRequest} interface, which allowed blocking tools to inspect and cancel network requests dynamically. Manifest V3 replaced this with a \texttt{declarativeNetRequest} interface, under which extensions submit rule sets that the browser engine evaluates according to its own logic, with constraints on the number and complexity of rules~\citep{cyphersGooglesManifestV32021}. The stated rationale invoked security (constraining the broad capabilities the old API granted to all extensions, including malicious ones) and performance (the declarative model permits engine-level optimization that arbitrary script-based filtering does not), framings that have been both reproduced and contested in critical analyses~\citep{cyphersGooglesManifestV32021, hancockManifestV3Open2021}.

While the API constraints themselves are not in dispute, how much they reduced practical blocking capability has been debated among extension developers. The effect is less disputed on Firefox, which retained the older API and where uBlock Origin continues to operate with full functionality, and on Brave, a Chromium-derived browser built to apply the same community filter lists directly rather than through an extension~\citep{braveShields}. Escape, however, now requires switching browsers.  The Coalition for Better Ads case and the Manifest V3 transition represent two distinct kinds of change in the ad-blocking landscape; whether one paved the way for the other, or whether they are better understood as independent developments, is a matter on which analysts have reached different conclusions.

\ifdefined\extendedversion

\subsection{Platform Recommender Systems}
\label{sec:case-recommenders}

Recommender systems organize what users encounter on large digital platforms: which videos appear in a YouTube feed, which products Amazon surfaces in search results, which posts Instagram shows in a timeline. Their basic architecture involves retrieving a candidate set from a very large corpus and then ranking candidates using a scoring model, with additional layers for safety, integrity, and content policy~\citep{cobbeRegulatingRecommendingMotivations2019, seaverCaptivatingAlgorithmsRecommender2018}. The two-stage design reflects the scale of contemporary platforms. By the 2010s catalogs on the largest platforms had grown to hundreds of millions of items; a single ranking model cannot evaluate the entire corpus per request~\citep{edelsonComparativeSurveyAlgorithmic2025, riederPlatformObservability2020}. These systems run on proprietary infrastructure with no external access to either the candidate retrieval logic or the ranking models~\citep{riederPlatformObservability2020}. Beyond the question of what any individual user sees, they govern the distribution of attention across entire media ecosystems: which creators are viable, which topics circulate, which commercial actors can reach audiences at scale.

Early recommender systems deployed on consumer platforms in the mid-2000s were oriented primarily toward relevance, returning content related to what a user had searched for or previously engaged with. YouTube’s early recommendation logic, for example, used click-through rate as a primary signal, directing users toward content that others with similar search behavior had chosen to watch~\citep{seaverCaptivatingAlgorithmsRecommender2018}. Amazon’s early collaborative filtering recommended products that users with similar purchase histories had bought, a logic that made the system’s workings relatively legible to the people it affected.

The most widely noted shift in YouTube’s history came around 2011–2012, when the platform moved from click-through rate to watch time and session length as its primary optimization targets~\citep{seaverCaptivatingAlgorithmsRecommender2018, xiangYouTubeProtocologicalControl2022}. Two accounts of this shift coexist in the literature and in YouTube’s own public statements, and they are not mutually exclusive. The first is a user-quality account: click-through rate rewarded misleading thumbnails and titles that generated clicks without delivering on their promise, whereas watch time was a more reliable signal that viewers found content worth completing. The second is a revenue account: watch time and session length convert directly to advertising impressions in a way that click-through rate, measured per video rather than per session, does not~\citep{cobbeRegulatingRecommendingMotivations2019}. The shift coincided with a period in which YouTube was working to make its advertising model viable at scale. Both accounts can describe what happened simultaneously; they imply different assessments of whose interests the new optimization target primarily served. The shift sat within a broader technical convergence on behavioral data as the basis for inferring preference at scale. By that point the recommender systems literature had established that implicit behavioral signals carry information about preference that explicit ratings or click-throughs do not capture: they are generated at higher volume, do not suffer from rating-scale ambiguity, and reflect what users actually do rather than what they say~\citep{huCollaborativeFilteringImplicit2008}. YouTube’s own engineering account, published in 2016, documents the operationalization of these principles in a two-stage deep neural network pipeline trained on watch behavior~\citep{covingtonDeepNeuralNetworks2016}. Whether watch time is in fact a better proxy for user satisfaction, or whether it systematically promotes content that is psychologically engaging but not valued in retrospect, has been a persistent and unresolved question in subsequent research~\citep{anwarRecommendationTemptation2025}.

Similar transitions occurred on other platforms. Instagram introduced an algorithmic feed in 2016, replacing reverse-chronological ordering with engagement-based ranking, and subsequently prioritized Reels, its short-form video product, in feed and discovery surfaces~\citep{singhAlgorithmicBiasSocial2023}. TikTok built its core product around an interest-graph model, using engagement signals from each viewing session to construct individualized feeds independent of social connections, a design that proved highly effective at generating prolonged sessions~\citep{wangRecommendationAlgorithmTikTok2022}. Across platforms, the architecture converged on behavioral signals (dwell time, completion rate, shares, comments) as the primary basis for inferring user preference. The individual outcome at stake is broader than relevance: personalization narrows what each user is shown, and the resulting filter bubbles are not observable from within the interface that produces them~\citep{pariserFilterBubble2011}. The infrastructure required to process these signals at scale is, by construction, accessible only to the platforms that own the data.

Empirical audits of contemporary recommender systems have documented systematic effects shaped by platform commercial interests. Amazon’s recommender systems favor the platform’s own private-label products over comparable third-party offerings, a pattern shaped by Amazon’s dual role as marketplace operator and seller on the same platform~\citep{dashWhenUmpireAlso2021}. On YouTube, integrity layers introduced under regulatory pressure have been reported to be designed in ways that limit their effect on watch time~\citep{krugerHaveYourCake2023}. Multi-objective approaches to recommender design, combining engagement with user-stated values, safety, and other objectives, have been proposed in the technical literature and partially implemented in industry integrity layers, with ongoing debate about whether such configurations meaningfully alter platform-level outcomes~\citep{strayBuildingHumanValues2024, krugerHaveYourCake2023}. External audits face significant limits when the underlying retrieval and ranking logic is proprietary and the platform controls what experimental access is available~\citep{messmerAuditingRecommenderSystems2023, fabbriAuditingRecommenderSystems2025}.

Regulatory attention began to accumulate in Europe in the early 2020s. The EU’s Digital Services Act (DSA), adopted in 2022, requires very large online platforms to document ranking criteria, conduct risk assessments for recommender systems’ societal effects, and offer users at least one recommendation option not based on behavioral profiling~\citep{reviglioGoverningPlatformRecommender2023, fabbriSelfdeterminationExplanationEthical2023}. The Digital Markets Act (DMA), also adopted in 2022, targets the largest platforms directly with requirements for interoperability and prohibitions on self-preferencing. Early audits of DSA compliance have found limited implementation of the mandated user-choice provisions and significant gaps between documented ranking criteria and what external researchers can observe about recommendation behavior~\citep{fabbriAuditingRecommenderSystems2025, messmerAuditingRecommenderSystems2023}. Whether the regulatory framework will produce changes beyond initial compliance, and whether the audit tools it creates can access the underlying systems, remains an open question.
\else
\input{content/case-recommenders-short}
\fi

\subsection{Financial Robo-Advisors}
\label{sec:case-roboadvisors}

Robo-advisors are digital platforms that provide automated investment advice and portfolio management to retail clients, typically through a web or mobile interface with minimal human adviser involvement~\citep{grealishRoboAdvisoryInvestingPrinciples2021, ruhrClassificationDecisionAutomation2019}. The basic pipeline is standardized: a client completes a questionnaire about their financial situation, investment goals, time horizon, and self-assessed risk tolerance; an algorithm classifies the client into a risk category; that category maps to a model portfolio implemented through exchange-traded funds (ETFs); and ongoing automation handles rebalancing, contributions, and, on some US platforms, tax-loss harvesting~\citep{grealishRoboAdvisorsTodayTomorrow2021, grealishRoboAdvisoryInvestingPrinciples2021}. Underneath sit modern portfolio theory and long-run expected-utility reasoning, which for any specified risk preference yield a small set of efficient allocations across broad asset classes~\citep{hayesActiveConstructionPassive2021, hayesEnactingRationalActor2020}. The questionnaire that begins the process performs two functions at once: it elicits information from the client, and through its sliders, projections, and default allocations, it also communicates back what kind of investor the client is~\citep{hayesActiveConstructionPassive2021, tanRoboadvisorsFinancializationLay2020, brauerNudgedBetterPortfolios2021}.

Commercial robo-advisors emerged in the US around 2008–2012 with firms such as Betterment and Wealthfront. For clients with small balances or no prior access to a human adviser, these platforms offered systematic diversification, low-cost ETF allocations, and automated services like rebalancing and lot-level tax-loss harvesting that would otherwise be difficult to perform manually~\citep{jungDesigningRoboadvisorRiskaverse2018, grealishRoboAdvisorsTodayTomorrow2021}. The platforms presented their technology as a neutral application of established financial theory made accessible at low cost through automation. The early design question was technical: how far personalization could extend beyond coarse risk categories without sacrificing scalability~\citep{faloonIndividualizationRoboAdvice2017}.

Digital advice developed inside an existing regulatory framework. Suitability requirements, “know your customer” processes, and fiduciary or best-interest standards in both the US and Europe long predate the technology and constrain the kinds of elicitation and recommendations that can be performed. In Europe, MiFID~II, PRIIPs, and the Insurance Distribution Directive specify how risk tolerance and capacity must be elicited, documented, and translated into recommendations; GDPR further regulates the use of personal data for automated profiling~\citep{steennotRoboadvisoryServicesInvestor2021, mrkyvkaRoboadvisoryDrivingForce2023, bertrandQuestioningAbilityFeaturebased2023}. In the US, SEC and FINRA rules play a similar role, with ERISA additionally governing retirement products~\citep{grealishRoboAdvisorsTodayTomorrow2021}. The short, structured questionnaires that became standard across the industry partly reflect this context: such formats produce documentation that supervisors can audit, backtest, and review against the firm’s stated suitability process~\citep{ruhrClassificationDecisionAutomation2019, tertiltAdviseNotAdvise2018}.

Between approximately 2018 and 2020, systematic audits of commercial platforms in the US and Europe produced a different picture of how the standardized pipeline was operating in practice. Risk questionnaires were often short, some questions had no observable effect on the resulting risk category, and platforms assigning the same risk label could end up with substantially different portfolios for the same client~\citep{tertiltAdviseNotAdvise2018, boreikoHowRiskProfiles2020}. The mapping from questionnaire responses to risk category, and from risk category to portfolio weights, was opaque; the heterogeneity across providers exceeded what differences in portfolio theory could account for~\citep{gasparRoboAdvisingInvestor2024}. Defaults dominated. Investors who had previously held all-equity portfolios stayed at the platform's default 50\% equity exposure, and first-time users were the least likely to move off it~\citep{brauerNudgedBetterPortfolios2021}. Qualitative work in the same period reframed robo-advisors as socio-technical assemblages that actively enact modern portfolio theory and Weberian rationality, “actively constructing passive investors” through automation, interface design, and behavioral defaults~\citep{hayesActiveConstructionPassive2021, hayesEnactingRationalActor2020, tanRoboadvisorsFinancializationLay2020}.

Many platforms, particularly those operated by or affiliated with large financial institutions, generate revenue not only from asset-based fees but from product-related sources: affiliated ETFs, internal fund platforms, and cash sweep programs that deposit uninvested balances in interest-bearing accounts at below-market rates~\citep{grealishRoboAdvisoryInvestingPrinciples2021}. These arrangements raised questions, in both the US and European regulatory contexts, about whether platforms operating in this way were meeting their legal obligations to act in clients’ best interests. Research on the implementation of explainability requirements under MiFID~II and GDPR found that feature-based explanations, introduced to meet legal expectations for transparency, sometimes increased user trust without improving user understanding of the underlying recommendation~\citep{bertrandQuestioningAbilityFeaturebased2023}. On this evidence, explainability functioned as a surface for building trust while leaving the underlying recommendation logic outside the feedback loop.

In the US, the SEC proposed extending conflict-of-interest rules to the algorithms themselves, then withdrew the proposal in 2025 after industry opposition, leaving the prior fiduciary framework in force~\citep{bearupQuickTakeSEC2023, secWithdrawPDA2025}.

\subsection{Spam Filters}
\label{sec:case-spam}

Spam governance has a longer history than the other cases examined here, and a richer critical literature that has itself contested how the history should be read. Brunton reads spam as a “shadow history” of the internet, one where fundamental questions about legitimate communication, commercialization, and identity have been fought out but rarely made visible as political questions~\citep{bruntonSpamShadowHistory2013}.

Email’s original protocol design made no provision for authenticating senders or distinguishing commercial bulk mail from other traffic~\citep{bandaySpamTechnologicalLegal2011}. Through the mid-1990s, as commercial use of email expanded rapidly, unsolicited bulk mail grew from a nuisance to an operational problem for administrators, who responded with ad hoc keyword filters and heuristic rules. The first coordinated response took the form of collaborative blocklists: DNS-based databases of known abusive IP addresses maintained by volunteer organizations, most notably Spamhaus and SpamCop, to which any mail server operator could subscribe~\citep{mathewRiskyBusinessSocial2017, bandaySpamTechnologicalLegal2011}. These systems functioned as a shared knowledge infrastructure: listing criteria were relatively transparent, delisting procedures existed and were contestable, and participation was open to any operator. Early critics documented cases where entire IP address ranges were listed, excluding regions or providers rather than specific abusers, a pattern that Lueg and Twidale called “mystery meat” filtering and that others characterized as “digital redlining”~\citep{luegMysteryMeatRevisited2007}. The debates about blocklist governance in this period covered who could list a sender, on what criteria, and what recourse a listed party had. They were conducted in public, in technical forums, and through legal challenges.

From the mid-2000s, spam operations industrialized. Large botnets running on compromised consumer machines replaced open relays as the primary sending infrastructure, and spam monetization consolidated around affiliate marketing programs for pharmaceutical sites, counterfeit goods, and other gray or illegal markets~\citep{levchenkoClickTrajectoriesEndToEnd2011, pitsillidisSpamValueChain2013}. Research tracing the spam value chain found that despite the apparent dispersion of sending infrastructure, revenue and operational control were concentrated at a relatively small number of monetization and payment nodes, the point in the chain the authors identified as most exposed to intervention~\citep{levchenkoClickTrajectoriesEndToEnd2011, raoEconomicsSpam2012}.

The defensive response to industrial-scale spam shifted from collaboratively maintained blocklists to machine-learning classifiers trained on large volumes of message data aggregated across multiple tenants~\citep{ferraraHistoryDigitalSpam2019, bandaySpamTechnologicalLegal2011}. Classifier-based approaches scaled better against high-volume polymorphic threats than the rule-based and blocklist approaches they displaced; as cross-tenant training data accumulated at large providers, the performance advantage grew~\citep{bandaySpamTechnologicalLegal2011, ferraraHistoryDigitalSpam2019}. These classifier-based systems operate on proprietary cross-tenant data, and they do not provide senders with information about why a given message is or is not delivered~\citep{ferraraHistoryDigitalSpam2019, carmiMediaDistortionsUnderstanding2020}. During the same period, sender authentication standards (the Sender Policy Framework (SPF), Sender ID, and later DomainKeys Identified Mail (DKIM)) were developed and gradually adopted, establishing technical mechanisms to verify that a given host was authorized to send on behalf of a domain and thereby creating a new basis for sender reputation~\citep{bandaySpamTechnologicalLegal2011}. Legal frameworks, including CAN-SPAM in the US and e-privacy directives in Europe, formalized a distinction between lawful commercial email and unlawful spam, though researchers consistently noted that legal frameworks lagged technical change and could not resolve the underlying economic dynamics~\citep{raoEconomicsSpam2012, bandaySpamTechnologicalLegal2011}.

The governance landscape underwent a parallel transformation. Mathew and Cheshire’s ethnographic study of the anti-spam community documents a transition from a small group of administrators coordinating informally, through mailing lists, conferences, and direct contact with network operators, to a more institutionalized landscape of blocklists with formal procedures, industry forums, and best-practice bodies~\citep{mathewRiskyBusinessSocial2017}. M3AAWG (the Messaging, Malware and Mobile Anti-Abuse Working Group) became the primary venue for industry coordination on deliverability standards and anti-abuse best practices, but its membership is gated, and its proceedings are not public. The criteria determining whether a given sender can reach inboxes are set in a forum that the senders themselves cannot access.

Consolidation accelerated in the early 2010s as Gmail, Outlook, and Yahoo expanded to handle a substantial fraction of global email volume~\citep{ferraraHistoryDigitalSpam2019}. These large providers possessed data volumes sufficient to train filtering systems far more capable than anything available to smaller operators, and their filtering decisions effectively set de facto standards for deliverability across the ecosystem. DMARC (Domain-based Message Authentication, Reporting, and Conformance), standardized in 2012, extended the authentication framework to allow domain owners to specify policies for how receivers should handle messages that fail SPF or DKIM checks, providing domain owners with new leverage over how their sending infrastructure was represented but also further centralizing deliverability judgment at large providers who could implement and honor DMARC policies consistently~\citep{bandaySpamTechnologicalLegal2011}.

Filtering effectiveness for ordinary users has improved substantially: inbox spam rates at major providers are much lower than they were in the mid-2000s~\citep{ferraraHistoryDigitalSpam2019}. At the same time, smaller and independent senders face significant and often opaque barriers to deliverability. The GDPR has constrained how cross-service reputational data can be shared and retained, creating tension between the data-intensive methods that dominate modern filtering and the legal requirements of data minimization~\citep{ferraraHistoryDigitalSpam2019}. The most resourced abuse actors have adapted to the authentication and filtering environment, shifting from high-volume bulk spam to targeted phishing and business email compromise, which requires different and more computationally intensive detection approaches and closer integration with broader security infrastructure~\citep{raoEconomicsSpam2012}. Access to that integrated security infrastructure requires the kind of organizational scale and industry relationships that smaller operators did not need in the blocklist era, when participation in the principal anti-spam coordination bodies was open to any operator.

\section{Depoliticization and User Empowerment}
\label{sec:analysis}

In each case the technology opened a space of contestable choices about infrastructure, knowledge, and objectives; in each case that space was progressively narrowed through structuring moves that embedded those choices as technical facts. Drawing on accounts of how governments place contested decisions at one remove from democratic challenge~\citep{burnhamNewLabourPolitics2001, burnhamDepoliticisationCommentBuller2006}, we call this pattern \emph{depoliticization}: contestable questions are relocated into technical or expert forms, where they can no longer be collectively challenged~\citep{flindersDepoliticisationPrinciplesTactics2006, mouffeReturnPolitical2005, ranciereDisagreementPoliticsPhilosophy1999}. The cases also show the reverse, \emph{politicization}. EasyList's community governance kept the question of what to block open to challenge, and the SEC's rulemaking on fiduciary obligations kept open whose interests an advisor must serve.

A second throughline runs alongside the depoliticization dynamic: individual outcomes and organized collective capacity move independently. In spam, individual inbox quality at major providers improved substantially as classification consolidated around proprietary systems, while the organized capacity to challenge filtering criteria collapsed in the same transition. This paper's central analytical contribution is identifying what drives the two apart.

\ifdefined\extendedversion

\begin{table}[t]
\small
\caption{The ad blocking case at two moments. Each dimension moved from a configuration users and list communities could contest to one settled by Google and the advertising industry. \Cref{sec:cases} traces how far the other trajectories followed the same path.}
\label{tab:adblocker-moments}
\begin{tabular}{@{} >{\raggedright\arraybackslash}p{0.15\columnwidth} >{\raggedright\arraybackslash}p{0.345\columnwidth} >{\raggedright\arraybackslash}p{0.345\columnwidth} @{}}
\toprule
& \textbf{Filter-list era (2006--2011)} & \textbf{Coalition standards and Manifest~V3 (2019--2024)} \\
\midrule
\emph{Infra\-structural} & \texttt{web\allowbreak Request} API: any extension may intercept and cancel requests & \texttt{declarative\allowbreak Net\allowbreak Request}: the engine evaluates capped rule sets; full capability only outside Chrome \\[4pt]
\emph{Epistemic} & Community filter lists: readable, forkable, open contribution & Coalition criteria enforced in-browser; the lists survive, but Chrome cannot run them in full \\[4pt]
\emph{Teleo\-logical} & What to block decided by users and list communities & ``Acceptable advertising'' defined by an industry coalition including the largest ad seller \\
\bottomrule
\end{tabular}
\end{table}

\Cref{tab:adblocker-moments} summarizes the shift for the ad blocking case.
\fi

\subsection{Intermediary Institutions Settling Contestation}
\label{sec:analysis:institutions}

One might expect the decisive contests over whom agents serve to occur between users and platforms directly. The ad blocker trajectory includes a period of direct technical competition: content delivery networks introduced obfuscated ad delivery, blockers refined selector syntax, and publishers deployed anti-blocking scripts~\citep{iqbalAdWarsRetrospective2017}. That contest did not produce the durable settlement. Across the trajectories, the structuring moves that shaped each outcome were often settled in intermediary institutions sitting between users and the actors who build and control technologies. These institutions exercise two functions that reinforce each other. They set the rules, specifying configuration choices that infrastructure actors then implement and thereby constraining what is technically possible at the layers above. And they frame those rules as settled, presenting choices as technical specifications or expert consensus rather than open questions. We call these functions \emph{material instantiation} and \emph{discursive legitimation}. Where principal-agent accounts locate the source of misalignment in the dyadic relationship between user and firm~\citep{dowdingAlgorithmicDecisionMakingPrincipal2024, duettingAlgorithmicContractTheory2024}, the cases show that the decisive settlements occurred at this institutional level, prior to and independent of any individual user's relationship with any particular agent.

The Coalition for Better Ads and M3AAWG illustrate how this works in practice. The Coalition produced concrete filtering criteria enforced through Chrome and framed the question of acceptable advertising as industry consensus, with governance that included no adversarial challenge mechanism and no public interest representation~\citep{megaliDigitalPlatformsMembers2022, blackburnDigitalPlatformsCircular2023}. M3AAWG operates on a similar logic: deliverability standards set through gated proceedings with non-public outputs, framed as expert consensus on abuse prevention, inaccessible to most of the senders whose practices they govern~\citep{mathewRiskyBusinessSocial2017}. In both cases the locus of contestation shifted from a publicly accessible forum to a closed institutional one. Organizing a challenge now requires standing in a body that was not designed to accommodate it.

The SEC's handling of fiduciary obligations and the EU's Digital Markets Act (DMA) and Digital Services Act (DSA) show the adversarial-enabling alternatives. A statutory public interest mandate, open comment periods, legal standing for public interest groups, judicial review, and revision authority counteract specific foreclosure mechanisms~\citep{bearupQuickTakeSEC2023}: for example, open comment keeps the commitments contestable; revision authority prevents material embedding from accumulating unchecked. The DMA targets the infrastructural dimension through interoperability mandates; the DSA targets the epistemic dimension through transparency requirements and algorithm-choice mandates, both enforced through adversarial proceedings~\citep{reviglioGoverningPlatformRecommender2023}. Regulatory capture remains possible, and resource disparities and revolving door dynamics are well documented~\citep{carpenterPreventingRegulatoryCapture2013}, but the architecture raises the cost of capture and keeps the value commitment visible as a political choice.

A growing critical literature has questioned whether transparency alone is adequate for algorithmic accountability~\citep{anannySeeingKnowingLimitations2018, pasqualeBlackBoxSociety2015}. Transparency reports a system's current configuration. The constitutive choices are the ones that produced that configuration, and they are often not part of what gets reported. Chrome's Manifest V3 documentation accurately describes which extension APIs are available; it does not reveal that the API design was a contestable choice among alternatives, or that it was made by an actor with commercial interests in the outcome. This is why transparency without the structural complements that the SEC model illustrates risks performing the discursive legitimation function of intermediary institutions; it makes the cage visible without providing a key.

The cases show that technical performance for individual welfare and collective political capacity can move in opposite directions, but not as a necessary tradeoff. Centralized spam filtering improved individual inbox quality; whether that improvement required eliminating organized challenge capacity is a causal question the cases do not settle. The ad blocker trajectory is more direct: uBlock Origin on Firefox, the community-governed path, blocks a far wider range of advertising than Chrome's enforcement of Coalition standards, which filters only the formats the Coalition designates unacceptable~\citep{megaliDigitalPlatformsMembers2022, cyphersGooglesManifestV32021}. The paper's analytical focus is on the location of political capacity over agents. The cases provide no support for the view that centralized control is a precondition for adequate technical performance.

\subsection{Conditions of Collective Contestation}
\label{sec:analysis:conditions}

Knowing where configurations get settled raises a further question: what material conditions determine whether organized communities can contest them? The cases reveal three conditions that jointly determine whether that capacity can form.

Ad blocking provides the positive instance. When filter lists were maintained through community governance with publicly readable syntax and open contribution, three conditions were met: the epistemic task of identifying ad-serving domains was performable with publicly observable information (public observability); the browser extension API gave users independent infrastructure through which community knowledge could be applied (independent infrastructure)~\citep{storeyFutureAdBlocking2017, snyderWhoFiltersFilters2020}; and EasyList's transparent rules and forkable structure kept the question of what to block visibly open to challenge (governance sustaining contestation). Manifest V3 degraded the infrastructure condition while leaving the knowledge base intact~\citep{cyphersGooglesManifestV32021}. That uBlock Origin retained full capability on Firefox, where the API remained intact, demonstrates which condition was operative: the same knowledge base and governance produced different outcomes depending on which browser provided the material basis for applying them.

Spam governance traces the collapse of the first two conditions. Early collaborative blocklists (Spamhaus, SpamCop, the Distributed Checksum Clearinghouse) satisfied them: the task was performable with publicly observable information, and decentralized email infrastructure provided the material basis for community knowledge to be applied~\citep{mathewRiskyBusinessSocial2017, bandaySpamTechnologicalLegal2011}. As spam industrialized, the task shifted to require behavioral data available only at large-provider scale~\citep{ferraraHistoryDigitalSpam2019}. Public observability was lost; independent infrastructure became insufficient without it. Individual inbox quality improved substantially over this period. The organized capacity to produce shared knowledge about filtering criteria and contest the standards governing them collapsed in the same transition. The governance condition failed with the other two. Spamhaus and its peers had settled filtering criteria in public exchange; the large providers that replaced them settled the same criteria inside their own classifiers. What arms-race accounts read as a technical victory for superior classifiers~\citep{nithyanandAdblockingCounterBlocking2016} was, from the standpoint of political capacity, a displacement of the material conditions through which organized challenge had been possible.

Recommender systems present the case where none of the conditions were ever present. Major platforms controlling recommender systems for content and commerce have controlled all three dimensions from their commercial launch onward: the infrastructural basis of the recommendation process, the behavioral data through which the epistemic task was performed, and the objectives the system optimized for. Because the choices were never publicly contested in the relevant sense, no neutral-appearing intermediary institution was ever established to settle them. The DSA transparency requirements have produced compliance documentation without constituting the organized constituency that would make those disclosures actionable~\citep{fabbriAuditingRecommenderSystems2025, messmerAuditingRecommenderSystems2023}. The market-selection account (that configurations reflect user preferences expressed through choice among alternatives) faces a prior difficulty here: users were never positioned to choose among meaningfully different recommendation configurations, because the infrastructure through which such differences could have existed was configured by the platform from the outset. Browser extensions that hide or re-sort feed items, and research prototypes that demonstrate alternative rankings, work on the list the platform has already assembled. They cannot see which candidates it considered or what its ranking was optimizing for.

The comparison yields three conditions for organized collective contestation to form and persist.

\begin{enumerate}
    \item \textbf{Public observability} for collective knowledge to \emph{form}: the epistemic task must be performable with information accessible without platform cooperation.
    \item \textbf{Independent infrastructure} for collective knowledge to be \emph{used productively}, operationalized through infrastructure the platform does not control.

    \item \textbf{Governance that sustains contestation} for collective knowledge to \emph{persist} under adversarial pressure: governance structures without a formal challenge pathway absorb rather than sustain organized capacity to contest what agents do.
\end{enumerate}

Whether rules formally permit challenge is a secondary question to whether the material conditions are in place.

\subsection{Durability and the Burden of Re-politicization}
\label{sec:analysis:lockin}

These conditions, once lost, have proven hard to restore. Across the trajectories, the dominant arrangements did not merely outperform alternatives on the relevant technical task~\citep{nithyanandAdblockingCounterBlocking2016, iqbalAdWarsRetrospective2017}: they displaced the concrete arrangements through which alternatives had been organized. When Google redesigned Chrome's extension API, it removed the infrastructure through which community filter lists had operated. The lists survived; Chrome could no longer run them in full. When spam classification consolidated around cross-tenant behavioral data, the publicly observable signals on which decentralized alternatives had depended were no longer sufficient, and the collaborative blocklist infrastructure lost the material basis it required. In both cases there was nothing weakened to reactivate because consolidating the dominant arrangement itself removed the conditions an alternative would have needed.

Foreclosure is durable because embedding is concrete: API specifications, data access regimes, institutional standards. Each dimension of contestable choice has two layers, the dimension's purpose and its implementation in specific artifacts and procedures. The two functions named in the preceding section, \emph{material instantiation} and \emph{discursive legitimation}, operate at this implementation layer. Whoever holds that layer decides the question, whatever the formal rules say. Google capped how many filtering rules an extension may load and settled what users may block; the large providers moved classification onto data only they held and settled what filters can know. In ad blocking and spam, consolidation displaced working alternatives at this layer. Recommender systems never had alternatives here, because platforms held the layer from the moment the systems launched.

Single-dimensional foreclosure leaves cross-dimensional leverage. Manifest V3 constrains the infrastructural dimension, but the value question of what users should be able to block remains open, and Firefox preserves a site from which it can be pursued. When foreclosure accumulates across dimensions simultaneously, those footholds are removed together. Chrome's integration of Coalition for Better Ads standards embedded a value commitment within an institutional arrangement and fused both with browser infrastructure in a single sequence. Challenging the value standard requires overcoming the infrastructural embedding; the Coalition's governance structure provides no institutional pathway for either. Spam shows the same compounding: proprietary filtering, engagement-based reputation scoring, and centralized infrastructure combine such that engaging any dimension requires engaging the others~\citep{ferraraHistoryDigitalSpam2019, mathewRiskyBusinessSocial2017}.

The concreteness of embedding also explains the limits of regulatory intervention. When institutions built for adversarial challenge succeed in politicizing a dimension, actors with enough material control move the contestable choice down to an implementation level the intervention does not reach, a pattern we call nested re-embedding. MiFID~II forced risk-profiling criteria into public view and created a governance structure for ongoing challenge~\citep{steennotRoboadvisoryServicesInvestor2021}; platforms responded by encoding commercial discretion in questionnaire design, default fund allocation, and profiling conventions that formally satisfy the requirements while preserving opacity at the implementation level~\citep{boreikoHowRiskProfiles2020, tertiltAdviseNotAdvise2018}. Early DSA compliance patterns in recommender systems show the same dynamic, with transparency requirements producing ranking documentation that audits have struggled to reconcile with observed system behavior~\citep{fabbriAuditingRecommenderSystems2025, messmerAuditingRecommenderSystems2023}. The pattern extends to litigation; in the consolidated suits over social media harm to minors, the court allowed design claims about parental controls and session limits to proceed and dismissed every claim that would have required changing how platforms rank and display content~\citep{inReSocialMediaMDL2023}.

The shift in threshold for political challenge follows the same logic. Contesting a filtering decision in the EasyList period meant proposing a change, forking the list, arguing in community forums. After the same question was embedded in Chrome's Better Ads enforcement, the path to changing what Chrome enforces requires demonstrating that ``acceptable advertising'' encodes a value commitment, identifying a challenge pathway the Coalition's governance does not provide, and overcoming material embedding in a browser with dominant market share~\citep{megaliDigitalPlatformsMembers2022}. That first step is what discursive legitimation costs a challenger. Once a standard circulates as expert consensus, anyone contesting it must first show that a choice was made at all. Switching to Firefox remains available. It changes one user's own configuration and leaves the standard governing the dominant browser in place. Access asymmetry reinforces the burden: Coalition governance was dominated by industry members, M3AAWG requires organizational resources most affected parties lack, and the Acceptable Ads Committee was funded by the actors it regulated~\citep{mathewRiskyBusinessSocial2017}. Foreclosure and the threshold for re-politicization accumulate together, through routine decisions that require no visible political act.

Adversarial institutional design prevents foreclosure only where it is coordinated with the material conditions identified in the preceding subsection. The account implies a falsifiability condition: trajectories where multi-dimensional foreclosure was readily reversed, or where sustained collective contestation persisted without those material conditions, would require it to be revised.

Depoliticization, across these trajectories, is a structural tendency in domains where the actors providing infrastructure have commercial interests in configuring what agents can do; this tendency drives the individual-collective divergence the opening named. The next section turns to the institutions now forming around agentic AI and asks whether they are built like the ones that sustained contestation or like the ones that foreclosed it.

\section{Looking Forward: Agentic AI}
\label{sec:agenticai}

\ifdefined\extendedversion

\begin{table*}[t]
\centering
\small
\caption{Agentic AI across the three dimensions. Column two applies the conditions of \cref{sec:analysis:conditions} to the current arrangements in MCP and the Agentic AI Foundation.}
\label{tab:agentic}
\begin{tabular}{@{}
    >{\raggedright\arraybackslash}p{0.13\linewidth}
    >{\raggedright\arraybackslash}p{0.40\linewidth}
    >{\raggedright\arraybackslash}p{0.40\linewidth}
    @{}}
\toprule
& \textbf{Current state}
& \textbf{What sustained contestation looks like} \\
\midrule

\emph{Infrastructural}
    & MCP under the Agentic AI Foundation; capability boundaries fixed in
      the specification and its defaults; most agents are hosted, calling
      only the servers their provider permits; Governing Board seats by
      membership tier
    & \emph{Independent infrastructure}: self-run clients and servers stay
      a working alternative to hosted agents, open to smaller players, with
      regulatory backing if adoption alone does not suffice \\[5pt]

\emph{Epistemic}
    & Agent tool calls and data flows readable only by the provider running
      the agent; protocol design discussions publicly recorded; the
      foundation's budget and membership decisions carry no such obligation
    & \emph{Public observability}: agents auditable by design, with tool
      calls and data flows readable by outside parties, including public
      bodies; the foundation's budget and membership decisions published,
      as its design discussions already are \\[5pt]

\emph{Teleological}
    & Agents transact for users among merchants and platforms whose
      interests the transaction affects; no fiduciary equivalent; value
      commitments settled through product decisions
    & A duty to users that an affected party can invoke, comparable to
      advisors' fiduciary duty; challenges rest on the audit record, and
      defaults stay open to revision \\

\midrule

\multicolumn{3}{@{}p{0.95\linewidth}@{}}{%
    \emph{Across dimensions}, \emph{governance that sustains contestation}:
    formal standing for user and public-interest representatives, and an
    adversarial mechanism through which specification decisions can be
    challenged on user-harm grounds. The foundation currently offers the
    public process without the standing or the challenge mechanism.} \\

\bottomrule
\end{tabular}
\end{table*}

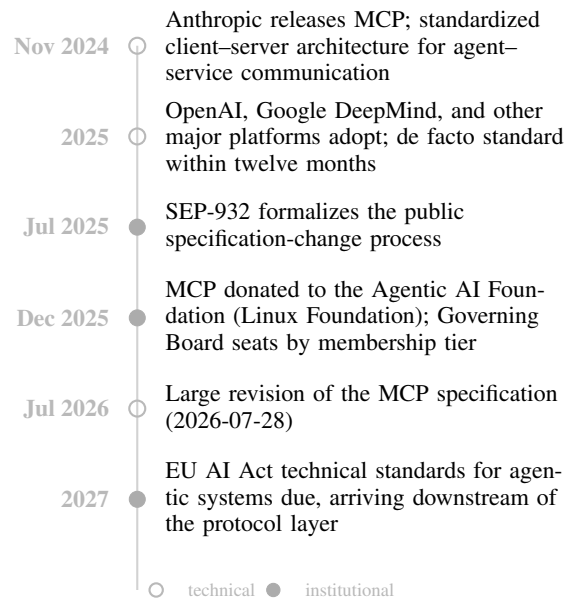
\begin{figure}[ht]
\centering
\begin{tikzpicture}
    \draw[gray!35, thick] (0, 0) -- (0, -7.2);
    \draw[gray!55, thick] (0, 0) circle (3pt);
    \node[anchor=east, font=\small\bfseries, text=gray!55] at (-0.25, 0) {Nov 2024};
    \node[anchor=west, font=\footnotesize, text width=5.4cm] at (0.25, 0) {Anthropic releases MCP; standardized client--server architecture for agent--service communication};
    \draw[gray!55, thick] (0, -1.2) circle (3pt);
    \node[anchor=east, font=\small\bfseries, text=gray!55] at (-0.25, -1.2) {2025};
    \node[anchor=west, font=\footnotesize, text width=5.4cm] at (0.25, -1.2) {OpenAI, Google DeepMind, and other major platforms adopt; de facto standard within twelve months};
    \filldraw[gray!65] (0, -2.4) circle (3pt);
    \node[anchor=east, font=\small\bfseries, text=gray!55] at (-0.25, -2.4) {Jul 2025};
    \node[anchor=west, font=\footnotesize, text width=5.4cm] at (0.25, -2.4) {SEP-932 formalizes the public specification-change process};
    \filldraw[gray!65] (0, -3.6) circle (3pt);
    \node[anchor=east, font=\small\bfseries, text=gray!55] at (-0.25, -3.6) {Dec 2025};
    \node[anchor=west, font=\footnotesize, text width=5.4cm] at (0.25, -3.6) {MCP donated to the Agentic AI Foundation (Linux Foundation); Governing Board seats by membership tier};
    \draw[gray!55, thick] (0, -4.8) circle (3pt);
    \node[anchor=east, font=\small\bfseries, text=gray!55] at (-0.25, -4.8) {Jul 2026};
    \node[anchor=west, font=\footnotesize, text width=5.4cm] at (0.25, -4.8) {Large revision of the MCP specification (2026-07-28)};
    \filldraw[gray!65] (0, -6.0) circle (3pt);
    \node[anchor=east, font=\small\bfseries, text=gray!55] at (-0.25, -6.0) {2027};
    \node[anchor=west, font=\footnotesize, text width=5.4cm] at (0.25, -6.0) {EU AI Act technical standards for agentic systems due, arriving downstream of the protocol layer};
    \draw[gray!55, thick] (0.25, -7.2) circle (2.5pt);
    \node[anchor=west, font=\scriptsize, text=gray!55] at (0.55, -7.2) {technical};
    \filldraw[gray!65] (1.8, -7.2) circle (2.5pt);
    \node[anchor=west, font=\scriptsize, text=gray!55] at (2.1, -7.2) {institutional};
\end{tikzpicture}
\caption{Agentic AI timeline to date. In the four historical trajectories (Appendix~\ref{app:timelines}), institutional consolidation followed years of community practice; here the protocol's institutional home, board composition, and governance process were settled within about two years of release, before any community-governance period and ahead of EU AI Act standards that will arrive downstream of the protocol layer.}
\label{fig:timeline-agentic}
\end{figure}

\fi

\paragraph{Shared structure, wider scope.} Each technology examined above acted on users' behalf in a domain where another party held an interest potentially opposed to the user's (\S\ref{sec:cases}). Agentic AI shares this characteristic: agents are poised to transact for their users among merchants, platforms, and services whose interests the transaction affects. It is also poised to act across domains, and potentially beyond those analyzed here. The structure is the same; the scope is wider. That widening means a single ecosystem could set the boundaries of possible agency across many domains at once. Because the structure is shared, the framework predicts where depoliticization will occur if this trajectory follows the others: in an industry-led intermediary that converts contestable choices into apparently technical specifications; through governance that is open at the surface while decision authority is held asymmetrically underneath; and without an adversarial mechanism to defend that openness once consolidation advances.

\paragraph{What is being settled.} Anthropic's Model Context Protocol (MCP), released in November 2024, defines how AI agents communicate with external services through a standardized client--server architecture~\citep{linuxFoundation2025AAIF}. Adoption by OpenAI, Google DeepMind, and other major platforms within months of release, together with the December 2025 donation to the Agentic AI Foundation (AAIF) under the Linux Foundation, established MCP as the de facto standard within twelve months~\citep{mcpBlog2025AAIFDonation}. The scope of what is settled at this layer is broad: what counts as a tool an agent may invoke, what data flows are permitted under what authorization, and how personal data is delineated when an agent acts on a user's behalf. Each is being fixed in the specification and its defaults, not in any statute.

The breadth is striking, and it covers the domains of our case studies. In each, agentic AI could unsettle what the earlier trajectory settled. Recall the introduction's examples: an agent building and maintaining a personal Chromium-based browser, or writing extensions that put the user's instructions ahead of the platform's preferences.

\paragraph{Where consequential choices are being settled.} On paper, the AAIF is open. The specification is public, anyone can submit changes through the Specification Enhancement Proposal (SEP) process, maintainer meetings are bi-weekly with public notes, and the Linux Foundation provides a neutral institutional home~\citep{sorriaParra2025MCPGovernance}. Binding authority sits elsewhere. Decision-making over strategic investments, budget, and membership criteria rests with a Governing Board composed of platinum members (AWS, Anthropic, Block, Bloomberg, Cloudflare, Google, Microsoft, OpenAI), with influence tied to financial commitment tier~\citep{linuxFoundation2025AAIF}. No formal standing exists for user organizations or public-interest representatives, no adversarial mechanism allows an affected party to contest an SEP outcome, and no public-interest mandate binds the Board. Individual users can configure their own permissions, and anyone can comment through the SEP process; the outcomes are decided by membership tier. Peer standards bodies show that stronger commitments are available to protocol governance: the IAB directs that protocol decisions favor end users where interests conflict~\citep{nottinghamInternetEndUsers2020}, and the W3C's design principles rank users' interests above those of page authors and implementors~\citep{vankesterenHTMLDesignPrinciples2007}. Nor is regulation currently positioned to reach these choices: the term ``agentic systems'' does not appear in the EU AI Act's legal text, so the technical standards under development for 2026--2027 will arrive downstream, operating within whatever the protocol layer has already settled~\citep{futureSociety2025AIAgentsEU}. This is the configuration \S\ref{sec:analysis:institutions} identified as discursive openness without structural protection: the Coalition for Better Ads pattern at a higher layer of infrastructure.

\paragraph{Is alignment enough?} The dominant approach to making agents serve their users is technical alignment: training and steering a model so that an individual agent acts in accordance with its user's values. The sophisticated end of this program takes value diversity seriously, asking how a model can represent many users' values rather than average over them~\citep{sorensenRoadmapPluralisticAlignment2024}. The trajectories traced here suggest that even this cannot fully secure the goal. Encoding values, plural or not, is work the provider performs on infrastructure the provider runs. And beyond any single agent, what counts as a tool an agent may invoke, which data flows are authorized, and which servers a hosted agent may call are settled in the specification and its governance before any particular agent is trained. Those settlements decide which alignments are achievable in the domain at all. Take the robo-advisor case: advice satisfied suitability requirements at each interaction while defaults and fund selection displaced user interests across the platform~\citep{boreikoHowRiskProfiles2020, tertiltAdviseNotAdvise2018}. Aligning an individual agent and keeping a domain contestable are distinct problems, and better training does not solve the second. Alignment that is to hold across providers and over time therefore depends on the conditions above: an audit record lets anyone verify that an agent served its user, and a governance channel lets an affected party contest a misaligned default rather than merely document it.

\paragraph{Preserving contestation.} Our analysis indicates that the sites where contestation can be lodged matter more than broad normative declarations. However, these sites have been rapidly solidifying: the AAIF has been standing up its governance since December 2025~\citep{linuxFoundation2025AAIF}, the MCP specification is being iterated upon, with an updated version released in July 2026~\citep{mcpSpec2026}, and the EU AI Act's technical standards for agentic systems are due for adoption in 2027~\citep{futureSociety2025AIAgentsEU}. This is a critical moment to establish formal standing for user and public-interest representatives, and an adversarial mechanism through which SEP outcomes can be challenged on user-harm grounds.

We now map to agentic AI the conditions that sustained contestation in the cases (\S\ref{sec:analysis:conditions}): public observability, independent infrastructure, and governance that sustains contestation.  Regarding public observability, the MCP project already records its decisions and the discussions behind them and makes them publicly available; the AAIF, whose budget and membership decisions carry no such obligation, would need to meet the same standard, and the agents themselves would need to be auditable, with tool calls and data flows readable by outside parties, including public bodies, rather than only by the provider running the agent. Regarding independent infrastructure, the specification is openly licensed and anyone can run their own client and servers today, but most agents are hosted by providers who decide which servers they may call, and keeping the protocol open to smaller players may require regulatory backing. Regarding governance, a duty to users comparable to the fiduciary duty the SEC imposed on advisors would give an affected party grounds to challenge a specification, and the audit record is what such a challenge would rest on.

Table~\ref{tab:agentic} sets the arrangements described above against the conditions the trajectories identify. The three dimensions are assessed separately because foreclosure on each proceeds through a different mechanism, while the governance condition is stated once: it is institutional rather than dimension-specific, and in its absence degradation on any dimension can proceed without a venue in which it can be contested.

\Cref{fig:timeline-agentic} collects the sequence: release, cross-platform adoption, governance formalization, and institutional consolidation within roughly two years, with the earliest regulatory instrument due in 2027. The case timelines in Appendix~\ref{app:timelines} give the comparison: EasyList ran for five years before Acceptable Ads, and the collaborative blocklists for eight before M3AAWG; here the institutional moves came first.

\section{Limitations and Conclusion}
\label{sec:discussion}

\paragraph{Limitations.} A key limitation concerns the goal of user empowerment itself.  We ask whether users will be empowered, not whether empowerment is beneficial. In some cases user empowerment is dangerous on its own terms: enabling a user to plan an attack, disrupt critical infrastructure, or target another person's life or property. Relatedly, empowerment can put users' interests in conflict with one another. The right to be forgotten illustrates this pattern: enacted as an individual right, it empowers the person seeking erasure at the expense of other parties' legitimate interests, including those who would search, those who would publish, and future users whose informational environment is being shaped. We do not address how to collectively empower users with diverse and conflicting interests, how to adjudicate between them, or how viable coalitions among users might be built.

The analysis is bounded in other ways as well. The cases examined are US‑ and EU‑centric and exclude other important global considerations.  In particular, the Chinese contributions to agentic AI already loom large.   
Moreover, the mechanisms through which constitutive politics operate in regulatory contexts without equivalent administrative law traditions or civil society infrastructure are not addressed and require separate analysis. 

At a methodological level, the three dimensions emerged from the trajectories we analyzed, and \S\ref{sec:framework} argues they are necessary for any computational agent; whether they are exhaustive, and whether additional dimensions become constitutive in particular domains, remains open. 
The nested re‑embedding dynamic (\S\ref{sec:analysis:lockin}) and the interactions among the three dimensions are characterized through the cases, not formally modeled. The account is open to revision: the falsifiability conditions stated in \S\ref{sec:analysis:lockin} specify the kinds of evidence that would require the mechanism claim to be reconsidered.  Additionally, the agentic AI section in \S\ref{sec:agenticai} is prospective, identifying structural conditions and the stakes of early decisions while the trajectories themselves have yet to unfold.

\paragraph{Conclusion.}  
We began by asking who wins the contest between users' agents and the platforms they act on. By our analysis, the decisive moves are made before play begins across three dimensions: teleological, infrastructural, and epistemic.  When institutions are able to convert political questions into technical settlements, users may benefit from an improved product, as spam was largely removed from inboxes, while losing the ability to affect the process and any future changes.  Agentic AI's rules are being written now and will simultaneously affect many domains. The key result is not individual outcomes but an ecosystem that allows vigorous ongoing contestation.  

\section*{Acknowledgments}
We thank the Political Economy and Algorithms Collective at the University of Michigan for their useful feedback. We also thank Ali Alkhatib for helpful feedback, and Professor David Temin at the University of Michigan for his feedback and suggestions on the political science approach in the project's early stages.

\section*{Disclosure Statement}
We used generative AI tools for editing and for searching sources, which we verified. Authors declare no competing interests.

\bibliography{agenticpower}

\clearpage
\appendix

\section{Case Timelines}
\label{app:timelines}

The following timelines mark the principal technical and institutional developments in each trajectory. Open circles indicate changes to infrastructure, protocols, or technical architecture; filled circles indicate governance, regulatory, or institutional moves.

\begin{figure}[ht]
\centering
\begin{tikzpicture}
    \draw[gray!35, thick] (0, 0) -- (0, -9.6);
    \draw[gray!55, thick] (0, 0) circle (3pt);
    \node[anchor=east, font=\small\bfseries, text=gray!55] at (-0.25, 0) {2006};
    \node[anchor=west, font=\footnotesize, text width=5.4cm] at (0.25, 0) {EasyList community filter list; open contribution process};
    \filldraw[gray!65] (0, -1.2) circle (3pt);
    \node[anchor=east, font=\small\bfseries, text=gray!55] at (-0.25, -1.2) {2011};
    \node[anchor=west, font=\footnotesize, text width=5.4cm] at (0.25, -1.2) {Acceptable Ads whitelist introduced (Eyeo)};
    \filldraw[gray!65] (0, -2.4) circle (3pt);
    \node[anchor=east, font=\small\bfseries, text=gray!55] at (-0.25, -2.4) {2014};
    \node[anchor=west, font=\footnotesize, text width=5.4cm] at (0.25, -2.4) {German courts uphold ad blocking as lawful};
    \draw[gray!55, thick] (0, -3.6) circle (3pt);
    \node[anchor=east, font=\small\bfseries, text=gray!55] at (-0.25, -3.6) {2015};
    \node[anchor=west, font=\footnotesize, text width=5.4cm] at (0.25, -3.6) {uBlock Origin released; no whitelisting program};
    \filldraw[gray!65] (0, -4.8) circle (3pt);
    \node[anchor=east, font=\small\bfseries, text=gray!55] at (-0.25, -4.8) {2017};
    \node[anchor=west, font=\footnotesize, text width=5.4cm] at (0.25, -4.8) {Coalition for Better Ads standards published};
    \filldraw[gray!65] (0, -6.0) circle (3pt);
    \node[anchor=east, font=\small\bfseries, text=gray!55] at (-0.25, -6.0) {2019};
    \node[anchor=west, font=\footnotesize, text width=5.4cm] at (0.25, -6.0) {Chrome enforces Coalition standards as built-in feature};
    \draw[gray!55, thick] (0, -7.2) circle (3pt);
    \node[anchor=east, font=\small\bfseries, text=gray!55] at (-0.25, -7.2) {2019};
    \node[anchor=west, font=\footnotesize, text width=5.4cm] at (0.25, -7.2) {Manifest V3 announced; declarativeNetRequest replaces webRequest};
    \draw[gray!55, thick] (0, -8.4) circle (3pt);
    \node[anchor=east, font=\small\bfseries, text=gray!55] at (-0.25, -8.4) {2024};
    \node[anchor=west, font=\footnotesize, text width=5.4cm] at (0.25, -8.4) {Manifest V3 enforced in Chrome; full functionality on Firefox only};
    \draw[gray!55, thick] (0.25, -9.6) circle (2.5pt);
    \node[anchor=west, font=\scriptsize, text=gray!55] at (0.55, -9.6) {technical};
    \filldraw[gray!65] (1.8, -9.6) circle (2.5pt);
    \node[anchor=west, font=\scriptsize, text=gray!55] at (2.1, -9.6) {institutional};
\end{tikzpicture}
\caption{Ad blocker trajectory. The 2011 Acceptable Ads program and 2017 Coalition for Better Ads are the principal institutional structuring moves; Manifest V3 is the infrastructural one. uBlock Origin on Firefox represents the countercurrent maintained outside both.}
\label{fig:timeline-adblockers}
\end{figure}
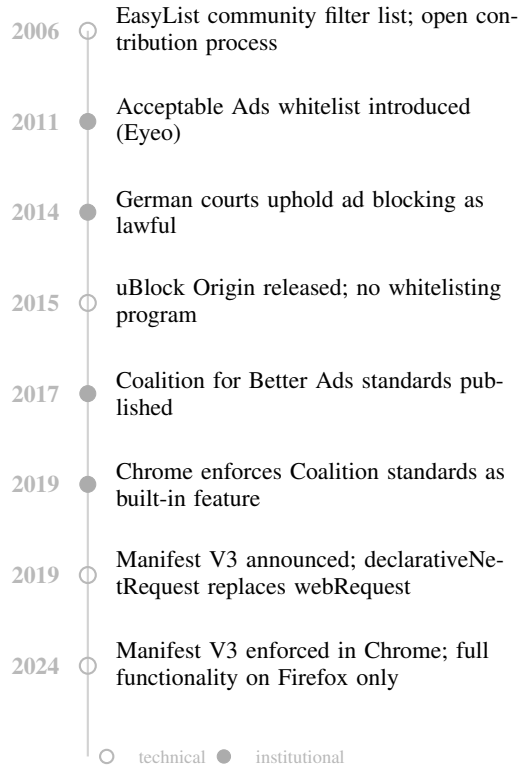

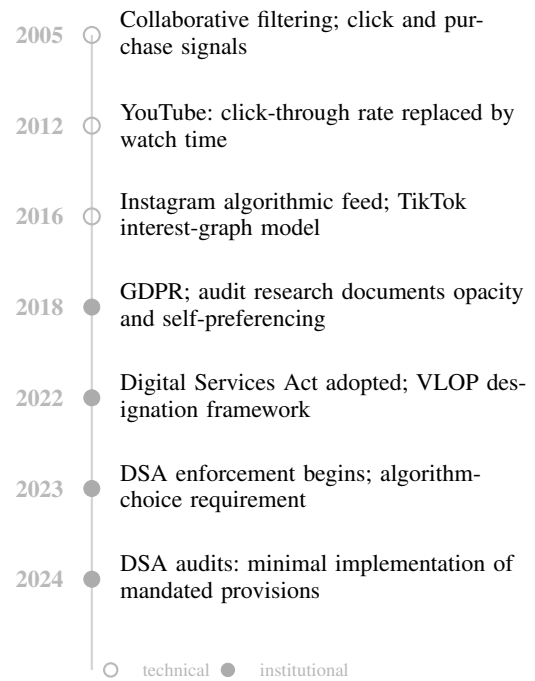
\begin{figure}[ht]
\centering
\begin{tikzpicture}
    \draw[gray!35, thick] (0, 0) -- (0, -8.4);
    \draw[gray!55, thick] (0, 0) circle (3pt);
    \node[anchor=east, font=\small\bfseries, text=gray!55] at (-0.25, 0) {2005};
    \node[anchor=west, font=\footnotesize, text width=5.4cm] at (0.25, 0) {Collaborative filtering; click and purchase signals};
    \draw[gray!55, thick] (0, -1.2) circle (3pt);
    \node[anchor=east, font=\small\bfseries, text=gray!55] at (-0.25, -1.2) {2012};
    \node[anchor=west, font=\footnotesize, text width=5.4cm] at (0.25, -1.2) {YouTube: click-through rate replaced by watch time};
    \draw[gray!55, thick] (0, -2.4) circle (3pt);
    \node[anchor=east, font=\small\bfseries, text=gray!55] at (-0.25, -2.4) {2016};
    \node[anchor=west, font=\footnotesize, text width=5.4cm] at (0.25, -2.4) {Instagram algorithmic feed; TikTok interest-graph model};
    \filldraw[gray!65] (0, -3.6) circle (3pt);
    \node[anchor=east, font=\small\bfseries, text=gray!55] at (-0.25, -3.6) {2018};
    \node[anchor=west, font=\footnotesize, text width=5.4cm] at (0.25, -3.6) {GDPR; audit research documents opacity and self-preferencing};
    \filldraw[gray!65] (0, -4.8) circle (3pt);
    \node[anchor=east, font=\small\bfseries, text=gray!55] at (-0.25, -4.8) {2022};
    \node[anchor=west, font=\footnotesize, text width=5.4cm] at (0.25, -4.8) {Digital Services Act adopted; VLOP designation framework};
    \filldraw[gray!65] (0, -6.0) circle (3pt);
    \node[anchor=east, font=\small\bfseries, text=gray!55] at (-0.25, -6.0) {2023};
    \node[anchor=west, font=\footnotesize, text width=5.4cm] at (0.25, -6.0) {DSA enforcement begins; algorithm-choice requirement};
    \filldraw[gray!65] (0, -7.2) circle (3pt);
    \node[anchor=east, font=\small\bfseries, text=gray!55] at (-0.25, -7.2) {2024};
    \node[anchor=west, font=\footnotesize, text width=5.4cm] at (0.25, -7.2) {DSA audits: minimal implementation of mandated provisions};
    \draw[gray!55, thick] (0.25, -8.4) circle (2.5pt);
    \node[anchor=west, font=\scriptsize, text=gray!55] at (0.55, -8.4) {technical};
    \filldraw[gray!65] (1.8, -8.4) circle (2.5pt);
    \node[anchor=west, font=\scriptsize, text=gray!55] at (2.1, -8.4) {institutional};
\end{tikzpicture}
\caption{Recommender systems trajectory. Unlike the other three cases, no community-governed period precedes commercial consolidation. Regulatory intervention via the DSA began in 2023; early audits document limited effect on underlying recommendation logic.}
\label{fig:timeline-recommenders}
\end{figure}

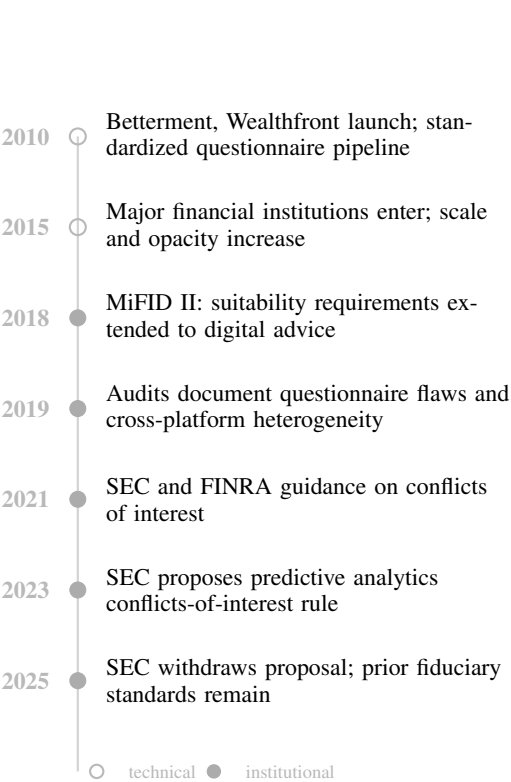
\begin{figure}[ht]
\centering
\begin{tikzpicture}
    \draw[gray!35, thick] (0, 0) -- (0, -8.4);
    \draw[gray!55, thick] (0, 0) circle (3pt);
    \node[anchor=east, font=\small\bfseries, text=gray!55] at (-0.25, 0) {2010};
    \node[anchor=west, font=\footnotesize, text width=5.4cm] at (0.25, 0) {Betterment, Wealthfront launch; standardized questionnaire pipeline};
    \draw[gray!55, thick] (0, -1.2) circle (3pt);
    \node[anchor=east, font=\small\bfseries, text=gray!55] at (-0.25, -1.2) {2015};
    \node[anchor=west, font=\footnotesize, text width=5.4cm] at (0.25, -1.2) {Major financial institutions enter; scale and opacity increase};
    \filldraw[gray!65] (0, -2.4) circle (3pt);
    \node[anchor=east, font=\small\bfseries, text=gray!55] at (-0.25, -2.4) {2018};
    \node[anchor=west, font=\footnotesize, text width=5.4cm] at (0.25, -2.4) {MiFID II: suitability requirements extended to digital advice};
    \filldraw[gray!65] (0, -3.6) circle (3pt);
    \node[anchor=east, font=\small\bfseries, text=gray!55] at (-0.25, -3.6) {2019};
    \node[anchor=west, font=\footnotesize, text width=5.4cm] at (0.25, -3.6) {Audits document questionnaire flaws and cross-platform heterogeneity};
    \filldraw[gray!65] (0, -4.8) circle (3pt);
    \node[anchor=east, font=\small\bfseries, text=gray!55] at (-0.25, -4.8) {2021};
    \node[anchor=west, font=\footnotesize, text width=5.4cm] at (0.25, -4.8) {SEC and FINRA guidance on conflicts of interest};
    \filldraw[gray!65] (0, -6.0) circle (3pt);
    \node[anchor=east, font=\small\bfseries, text=gray!55] at (-0.25, -6.0) {2023};
    \node[anchor=west, font=\footnotesize, text width=5.4cm] at (0.25, -6.0) {SEC proposes predictive analytics conflicts-of-interest rule};
    \filldraw[gray!65] (0, -7.2) circle (3pt);
    \node[anchor=east, font=\small\bfseries, text=gray!55] at (-0.25, -7.2) {2025};
    \node[anchor=west, font=\footnotesize, text width=5.4cm] at (0.25, -7.2) {SEC withdraws proposal; prior fiduciary standards remain};
    \draw[gray!55, thick] (0.25, -8.4) circle (2.5pt);
    \node[anchor=west, font=\scriptsize, text=gray!55] at (0.55, -8.4) {technical};
    \filldraw[gray!65] (1.8, -8.4) circle (2.5pt);
    \node[anchor=west, font=\scriptsize, text=gray!55] at (2.1, -8.4) {institutional};
\end{tikzpicture}
\caption{Robo-advisor trajectory. MiFID II in 2018 is the adversarial regulatory intervention the analysis identifies as the closest historical analogue to the structural features currently absent from AAIF governance. The 2025 SEC withdrawal illustrates the fragility of that intervention under changed administration.}
\label{fig:timeline-roboadvisors}
\end{figure}

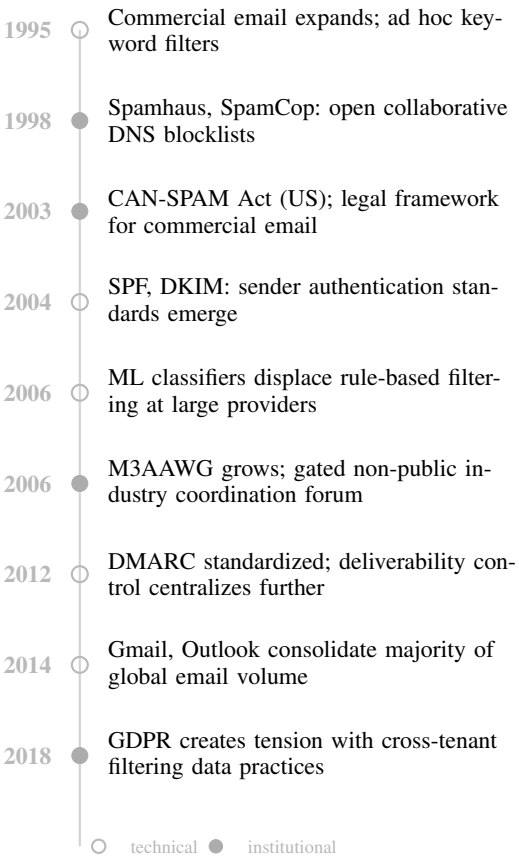
\begin{figure}[ht]
\centering
\begin{tikzpicture}
    \draw[gray!35, thick] (0, 0) -- (0, -10.8);
    \draw[gray!55, thick] (0, 0) circle (3pt);
    \node[anchor=east, font=\small\bfseries, text=gray!55] at (-0.25, 0) {1995};
    \node[anchor=west, font=\footnotesize, text width=5.4cm] at (0.25, 0) {Commercial email expands; ad hoc keyword filters};
    \filldraw[gray!65] (0, -1.2) circle (3pt);
    \node[anchor=east, font=\small\bfseries, text=gray!55] at (-0.25, -1.2) {1998};
    \node[anchor=west, font=\footnotesize, text width=5.4cm] at (0.25, -1.2) {Spamhaus, SpamCop: open collaborative DNS blocklists};
    \filldraw[gray!65] (0, -2.4) circle (3pt);
    \node[anchor=east, font=\small\bfseries, text=gray!55] at (-0.25, -2.4) {2003};
    \node[anchor=west, font=\footnotesize, text width=5.4cm] at (0.25, -2.4) {CAN-SPAM Act (US); legal framework for commercial email};
    \draw[gray!55, thick] (0, -3.6) circle (3pt);
    \node[anchor=east, font=\small\bfseries, text=gray!55] at (-0.25, -3.6) {2004};
    \node[anchor=west, font=\footnotesize, text width=5.4cm] at (0.25, -3.6) {SPF, DKIM: sender authentication standards emerge};
    \draw[gray!55, thick] (0, -4.8) circle (3pt);
    \node[anchor=east, font=\small\bfseries, text=gray!55] at (-0.25, -4.8) {2006};
    \node[anchor=west, font=\footnotesize, text width=5.4cm] at (0.25, -4.8) {ML classifiers displace rule-based filtering at large providers};
    \filldraw[gray!65] (0, -6.0) circle (3pt);
    \node[anchor=east, font=\small\bfseries, text=gray!55] at (-0.25, -6.0) {2006};
    \node[anchor=west, font=\footnotesize, text width=5.4cm] at (0.25, -6.0) {M3AAWG grows; gated non-public industry coordination forum};
    \draw[gray!55, thick] (0, -7.2) circle (3pt);
    \node[anchor=east, font=\small\bfseries, text=gray!55] at (-0.25, -7.2) {2012};
    \node[anchor=west, font=\footnotesize, text width=5.4cm] at (0.25, -7.2) {DMARC standardized; deliverability control centralizes further};
    \draw[gray!55, thick] (0, -8.4) circle (3pt);
    \node[anchor=east, font=\small\bfseries, text=gray!55] at (-0.25, -8.4) {2014};
    \node[anchor=west, font=\footnotesize, text width=5.4cm] at (0.25, -8.4) {Gmail, Outlook consolidate majority of global email volume};
    \filldraw[gray!65] (0, -9.6) circle (3pt);
    \node[anchor=east, font=\small\bfseries, text=gray!55] at (-0.25, -9.6) {2018};
    \node[anchor=west, font=\footnotesize, text width=5.4cm] at (0.25, -9.6) {GDPR creates tension with cross-tenant filtering data practices};
    \draw[gray!55, thick] (0.25, -10.8) circle (2.5pt);
    \node[anchor=west, font=\scriptsize, text=gray!55] at (0.55, -10.8) {technical};
    \filldraw[gray!65] (1.8, -10.8) circle (2.5pt);
    \node[anchor=west, font=\scriptsize, text=gray!55] at (2.1, -10.8) {institutional};
\end{tikzpicture}
\caption{Spam governance trajectory. The 1998 collaborative blocklists represent the epistemic commons whose displacement the analysis traces. The co-occurrence of the 2006 ML classifier shift and M3AAWG institutionalization marks the period in which technical scale advantage and governance access consolidated simultaneously in the same set of large providers.}
\label{fig:timeline-spam}
\end{figure}

\section{Conditions Across the Trajectories}
\label{app:conditions}

Table~\ref{tab:conditions} summarizes the status of the three conditions of \cref{sec:analysis:conditions} in each trajectory. The ad blocker, spam, and recommender rows restate the comparison that subsection draws. The robo-advisor row extends it to the case \cref{sec:analysis:institutions,sec:analysis:lockin} analyze through its institutions: there, MiFID~II created the observability the market had not, and the platforms' response moved commercial discretion into questionnaire design and defaults.

\begin{table*}[t]
\small
\caption{The three conditions of \cref{sec:analysis:conditions} in each trajectory. In spam, all three collapsed while inbox quality improved.}
\label{tab:conditions}
\begin{tabular}{@{} >{\raggedright\arraybackslash}p{0.11\textwidth} >{\raggedright\arraybackslash}p{0.19\textwidth} >{\raggedright\arraybackslash}p{0.19\textwidth} >{\raggedright\arraybackslash}p{0.19\textwidth} >{\raggedright\arraybackslash}p{0.19\textwidth} @{}}
\toprule
\textbf{Case} & \textbf{Public observability} & \textbf{Independent infrastructure} & \textbf{Governance sustaining contestation} & \textbf{Outcome for contestation} \\
\midrule
Ad blockers & Met: blocking targets identifiable from public page content & Met until Manifest~V3 removed the API in Chrome; intact on Firefox & Met: readable, forkable lists; displaced by Coalition enforcement & Sustained via uBlock Origin on Firefox; foreclosed in Chrome \\[4pt]
Recommenders & Never present: preference inference requires platform-held behavioral data & Never present: retrieval and ranking run on platform servers & Never present: no intermediary institution ever formed & Never formed; DSA disclosures without an organized constituency \\[4pt]
Robo-advisors & Created by regulation: MiFID~II forced profiling criteria into view & Never present: advice runs on broker and compliance infrastructure & SEC and MiFID~II rulemaking provide standing and revision authority & Re-opened by rulemaking, then re-embedded in questionnaire design and defaults \\[4pt]
Spam & Met, then lost when classification moved to provider-held behavioral data & Met, then insufficient once observable signals no longer decided delivery & Met: open blocklists; then gated (M3AAWG) & Collapsed while inbox quality improved \\
\bottomrule
\end{tabular}
\end{table*}

\section{Pattern of Depoliticization}
\label{app:analysis}

\begin{figure*}[t]
\centering
\begin{tikzpicture}[
    lab/.style={font=\scriptsize\itshape, text=gray!60, inner sep=1pt},
    evt/.style={font=\scriptsize, text=gray!55, inner sep=1pt},
    yr/.style={font=\scriptsize, text=gray!55},
]
    \fill[gray!6] (0,0)    rectangle (14.72,1.1);
    \fill[gray!6] (0,2.3)  rectangle (14.72,3.4);
    \node[lab, anchor=east, align=right, text width=1.5cm] at (-0.25,2.85) {all three\\met};
    \node[lab, anchor=east, align=right, text width=1.5cm] at (-0.25,1.7)  {some met};
    \node[lab, anchor=east, align=right, text width=1.5cm] at (-0.25,0.55) {none met};

    \draw[gray!40] (0,-0.15) -- (14.72,-0.15);
    \foreach \x/\y in {1.84/2000, 4.14/2005, 6.44/2010, 8.74/2015, 11.04/2020, 13.34/2025}{
        \draw[gray!40] (\x,-0.15) -- (\x,-0.25);
        \node[yr, anchor=north] at (\x,-0.28) {\y};
    }

    \draw[gray!75, line width=1pt, rounded corners=3pt]
        (0.92,3.1) -- (4.6,3.1) -- (6.44,0.75) -- (14.5,0.75);
    \node[lab, anchor=south west] at (0.85,3.18) {spam: collaborative blocklists};
    \node[evt, anchor=west, align=left] at (6.6,1.7) {ML classifiers;\\M3AAWG (2006)};

    \draw[gray!75, line width=1pt, dashed, rounded corners=3pt]
        (4.6,2.85) -- (10.58,2.85) -- (13.11,1.75) -- (14.5,1.75);
    \draw[gray!60, line width=0.5pt, dashed]
        (10.58,2.85) -- (14.5,2.85);
    \node[lab, anchor=south west] at (5.3,2.92) {ad blockers: EasyList};
    \node[evt, anchor=east, align=right] at (10.45,2.3) {Coalition in Chrome;\\Manifest V3 (2019)};
    \node[lab, anchor=south east, align=right] at (14.5,2.95) {uBlock Origin on Firefox\\(small share)};
    \node[lab, anchor=north west] at (13.3,1.68) {Chrome};

    \draw[gray!70, line width=1pt, dotted]
        (4.14,0.45) -- (14.5,0.45);
    \node[lab, anchor=south west] at (4.1,0.52) {recommenders};
    \draw[gray!55, fill=white] (12.42,0.45) circle (2pt);
    \node[evt, anchor=north] at (12.42,0.28) {DSA (2023)};

    \draw[gray!75, line width=1pt, dash dot, rounded corners=3pt]
        (6.44,0.2) -- (10.12,0.2) -- (10.58,1.55) -- (11.5,1.55) -- (14.5,1.3);
    \node[lab, anchor=east] at (6.3,0.2) {robo-advisors};
    \node[evt, anchor=east] at (9.95,1.0) {MiFID~II (2018)};
    \node[evt, anchor=north] at (13.0,1.38) {re-embedding};

    \draw[gray!50, line width=0.8pt]
        (13.3,2.1) -- (14.08,2.1);
    \node[lab, anchor=south west] at (13.28,2.2) {agentic AI: MCP (2024)};
    \node[evt, anchor=west, text=gray!60] at (14.16,2.1) {?};

\end{tikzpicture}
\caption{Conditions for collective contestation over time, by case. Vertical position is the band of \cref{sec:analysis:conditions} conditions met, following the readings argued in \cref{sec:cases,sec:analysis}; it is not a measurement. Lines that fall do not recover: spam after the 2006 shift to provider-held classifiers, ad blocking in Chrome after 2019. The thin strand is uBlock Origin on Firefox, a single-digit share of browsing. MiFID~II lifted robo-advice in 2018 and re-embedding pulled it part-way back. The agentic AI line begins in 2024.}
\label{fig:trajectories}
\end{figure*}
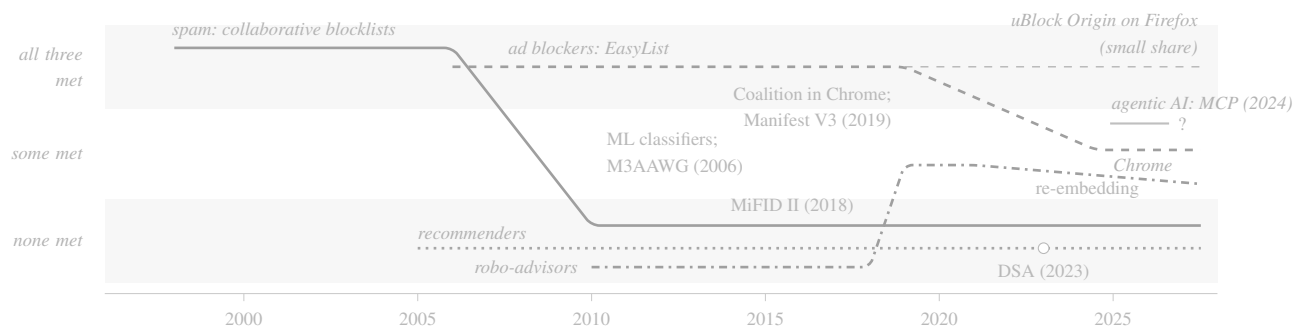

Figure~\ref{fig:trajectories} traces the conditions of \cref{sec:analysis:conditions} across the five trajectories; the falls mark the structuring moves \cref{sec:cases} dates.

\end{document}